\documentstyle[12pt]  {article}
\newcommand{\beq}{\begin{equation}}
\newcommand{\eeq}{\end{equation}}
\newcommand{\beqa}{\begin{eqnarray}}
\newcommand{\eeqa}{\end{eqnarray}}
\newcommand{\ba}{\begin{array}}
\newcommand{\ea}{\end{array}}

\begin{document}

\begin{center}
{\large \bf Ion Collisions in Very Strong Electric Fields}

\vskip 1.0 truecm

{\bf Luca Salasnich}  \\
Dipartimento di Fisica ``Galileo Galilei" dell'Universit\`a di Padova, \\
INFN, Sezione di Padova, \\
Via Marzolo 8, I 35131 Padova, Italy\\
and\\
Departamento de Fisica Atomica, Molecular y Nuclear, \\
Facultad de Ciencias Fisicas, Universidad ``Complutense" de Madrid, \\
Ciudad Universitaria, E 28040 Madrid, Spain \\

\vskip 0.5 truecm

{\bf Fabio Sattin}  \\
Dipartimento di Ingegneria Elettrica dell'Universit\`a di Padova, \\
Via Gradenigo 6/a, I 35131 Padova, Italy\\

\vskip 1.5 truecm
{\bf Abstract}
\end{center}

\vskip 0.5 truecm
\par
A Classical Trajectory Monte Carlo (CTMC) simulation has been made of
processes of charge exchange and ionization
between an hydrogen atom and fully stripped ions embedded in very strong
static electric fields ($O(10^{10}$ V/m$)$), which are thought to exist
in cosmic and laser--produced plasmas.
Calculations show that the presence of the field affects absolute values of
the cross sections, enhancing ionization and reducing charge exchange.
Moreover, the overall effect depends upon the relative orientation between
the field and the nuclear motion. Other features of a null-field situation,
such as scaling laws, are revisited.

\begin{center}
To be published in Journal of Physics B
\end{center}

\newpage

{\bf I. Introduction}
\vskip 0.5 truecm
\par
The study of the effects of externally imposed electric or magnetic fields
on atomic systems has been since a long time an active area of research in
physics, both when dealing with isolated atoms (Stark and Zeeman effects,
photoionization, chaotic spectra, modifications of energy levels and
wave, see e.g. Seipp and Taylor 1994, Wang and Greene 1994,
Delande and Gay 1986, Friedrich 1990, Zang and Rustgi 1994,
Delande {\it et al} 1994, Bylicki {\it et al} 1994, Buchleitner {\it et al}
1994), as when the attention is focused onto many--particle processes
(electron--atom or atom--atom collisions).
\par
In order to influence the electronic dynamics it is necessary that
the force exerted by the field is comparable with the Coulomb
intra--atomic forces, which are quite strong if the atom is not in a highly
excited state: for magnetic and electric fields the natural units of measure
are
$B_0 =  m_e^2 e^3 c / \hbar^3 = 2.35 \cdot 10^5$ T, and
$F_0 = m_e^2 e^5 / \hbar^4 = 5.14 \cdot 10^{11}$ V/m.
Fields of such strengths may be
generated only under rather exotic situations:
in white dwarfs magnetic fields of $ 10^5$ T, and in neutron
stars up to $10^9$ T, have been discovered (Ruder {\it et al} 1994).
The huge unipolar induction currents which generate the magnetic field
are sustained by differences of potential of order $O(10^{13}$ V/m$)$
(Ruder {\it et al} 1994).
By a different mechanism, electric fields may be generated even independently
by magnetic fields: it is the double layer (DL) phenomenon which is expected
to be found, for example, on the surface of neutron stars (Williams {\it et al}
1986, Raadu 1989).
A DL is a local discontinuity
surface consisting microscopically of two oppositely charged layers of plasma.
The resultant electric field within this region is very strong and acts to
re--establish a neutrality condition which is violated, for example,
as a consequence of temperature gradients on the surface of the star
or because of charge density gradients produced by the
different rates at which electrons and ions, falling
on the star surface from the outer space, are decelerated (Williams {\it et al}
 1986).
It is thought that fields $ > 10^{10}$ V/m may be generated through
this mechanism.
\par
DLs of considerable strength are created also in Earth laboratories during
laser experiments devoted, for example, to inertial confinement fusion
studies: in these
experiments a small target of hydrogen isotopes is irradiated with high power
($ \approx 10^{17} \div 10^{20}$ W/m$^2$) laser pulses. The sudden heating,
which converts into a hot plasma most of the target, drives a fast
dilatation of the gaseous outer corona where, because
of their higher velocity, the electrons will tend to lead the expansion.
The result is a DL which slows down the electrons and accelerates the ions
(for a review of the subject see Eliezer and Hora 1989).
Numerical simulations show that, inside these DLs which have a width
of about $10^{-7} \div 10^{-5}$ m, electric fields of $10^8 \div
10^{11}$ V/m may be reached (Eliezer and Hora 1989, Eliezer {\it et al} 1988).
\par
Charged particles within a DL usually have
great kinetic energies because of the heating, the electrostatic force,
and---in stars---the
gravitational attraction: they may be $O(10 $KeV/amu $)$
(Williams {\it et al} 1986, Eliezer and Hora 1989). DLs thus provide us
with rather extreme conditions where to examine interatomic processes,
particularly collisions. In this paper we aim to perform a numerical
investigation of the processes of electron capture and ionization between a
neutral atom and a bare ion. This kind of study is not new in literature:
 a similar work has been performed by Grosdanov and McDowell (1985) and
McDowell and Zardona (1985) for very strong, astrophysical
magnetic fields. Electric fields have been instead studied by Olson and
MacKellar (1981). There, however, the attention was focused on fields
attainable in ordinary laboratory
conditions, i.e. much weaker than those considered here; to compensate for the
smallness of the field only Rydberg atoms had to be considered. Here we
carry on the analysis on low lying electronic states instead that highly
excited ones; some topics not covered by Olson and MacKellar (1981) are
touched: the results of the
scattering process, expressed in terms of cross sections, are studied with
respect to the charge of the incident particle and the initial state of
the system. Also, particular attention is paid to the geometrical features
(direction and spatial extension) of the external field.

\vskip 0.5 truecm

{\bf II. Theory}
\vskip 0.3 truecm
\par
The collision process has been studied using the CTMC method, in which both
nuclei and the electron are
considered classical particles obeying Newton laws but the initial
conditions of the electron are randomly chosen within a distribution
which simulates the quantum mechanical behavior.
The study of the process is then reduced
to numerically integrate a set of coupled ordinary differential equations.
\par
The CTMC method, developed by Abrines and Percival (1966a), has been
frequently used
in studies of atomic scattering, being quite easy to implement and rather
accurate in results. The method works at its best in the region of
intermediate impact energies, i.e. when the relative impact internuclear speed
$v_p$ is greater than the classical orbital electron velocity $v_e$:
$ v_p \geq v_e$; in this work we will consider the range
$ 1 \leq v_p/v_e \leq 3.5$
(corresponding to energies $ E \leq 300$ KeV/amu), which  is quite in
accordance with the velocities expected to be found in DLs.
\par
The equations of motion for each of the particles are derived from the
Hamiltonian
\beq
 H = \sum_{i=1}^3 {p_i^2 \over 2}  +  \sum_{i<j =1}^3
{Z_i Z_j \over|{\vec q_i} -{\vec q_j}| } +
          \sum_{i=1}^3 {Z_i{\vec q_i}}  \cdot  {\vec F}\; ,
\eeq
where ${\vec q_i}$, ${\vec p_i}$ ($i = 1,2,3$) are the coordinates and the
momenta conjugate for the two nuclei and the electron;
$Z_i$ the charge of the i--{\it th} particle, and ${\vec F}$
the electric field (here and in the following, for the sake of easiness,
the electric field will be expressed in atomic units; in these units $F_0$
of section one is equal to 1).
\par
An atom embedded in a strong electric field is an instable system:
classically, a bound electron is field--ionized if the strength $F$ of
the field is greater than $E_b^2/4$, with $E_b$ binding energy. Thus, any
classical calculation loses meanings at strengths beyond this value
(for an hydrogen atom in the ground state, $E_b^2/4 = F_0/16 = 0.0625$ au).
However, quantum mechanically,
there is not a ionization threshold, instead all bound states
turn into resonances of finite lifetime, whatever small $F$ be. Several
calculations of lifetimes exist in literature; here we shall refer to
Damburg and Kolosov (1976):
in that work it is estimated that, in a field of $F=0.04$ au the mean life
of an electron in the ground state is
$\tau_e \approx 2 \cdot 10^{-11}$ s.
A particle moving at a speed $v_p \geq 10^6$ m/s (corresponding to
a kinetic energy $\geq 25$ KeV/amu) travels during this time a distance
$l_p = v_p \tau_e \geq 10^{-5}$ m. If we refer, for example, to a
laser--produced DL, its characteristic spatial length is given by the Debye
length $\lambda_D$; usually, this value is $\le 10^{-5}$ m
(see Eliezer and Hora 1989, sect. 4).
\par
Further, in a laser--produced DL the electric field is not at all static,
the timescale needed to establish it and over which its variations are not
negligible being in the range  of $10^{-13} \div 10^{-12} $ s
(Eliezer and Hora 1989, sects. 4,5).
Even within
such short times, a particle at the speed $v_p$ travels for several
hundreds of $\AA$, which is well beyond the typical length over which charge
transfer processes take place. So we see that, even under rather extreme
conditions, a classical description of the collision process does not
lose its meaning (we remark however that for cosmic DLs the picture is not so
satisfactory, because the width of the DL is supposed to be greater, thus in
this case the validity of our hypotheses may be questionable).
\par
Two further remarks need to be made when dealing with
excited atoms (we will consider only $n=2$): first, the validity of
the classical approximation still holds because we shall consider weaker
fields,
{\it ad hoc } rescaled to compensate for the smaller binding energy.
Further, because of the Stark mixing between levels due to $F$, and in
particular between the levels $2s - 2p_0$,
all the electrons are allowed to decay by spontaneous emission of radiation
into the ground state within short times ($\approx 10^{-9}$ s) which, however,
are always longer that the timescales of the collision process.
\par
The electric field exerts also an influence on electronic energies
(Stark shift) and wave functions. As far as the first are concerned, we see
from Damburg and Kolosov (1976) that the correction to the null-field values
is about 1 percent and thus may be neglected in the following.
\par
Using a bit more care instead is needed when dealing with position and momentum
distributions; in CTMC method quantum wave functions are replaced with a random
sampling of the initial coordinates from statistical distributions. The older
and more used choice is the microcanonical distribution (Abrines and Percival
1966b), where
momentum $\vec p_e$ and position $\vec q_e$ are picked up from an uniform
distribution subject to the bound that the total energy is equal to its
quantum mechanical value,
\beq
E = { |\vec p_e|^2 \over 2} - { 1 \over |\vec q_e|} = - { 1 \over 2 n^2 }
\eeq
This method provides a statistical momentum distribution which is equal
to the quantum mechanical one, while the  radial distribution is rather bad,
showing a cut--off beyond $r = |{\vec q_e}|= 2 n^2$ au .
This causes a wrong estimate of total cross section at low impact energies,
underestimating the contribution from collisions at large impact parameters.
Several methods have been devised to overwhelm this deficiency, with
good results
(see Hardie and Olson 1983, Cohen 1985, and also Eichenauer {\it et al} 1981,
Kunc 1988, Reinhold and Falcon 1988a,b, Schmidt {\it et al} 1990 for a
general discussion over the CTMC and other classical approximations).
However, any classical method which attempts to simulate a correct radial
quantum distribution has to give up the assumption that electron energy
is  a well fixed quantity. The electron will thus have some
probability to be given an energy beyond the
ionization threshold when the field is turned on, and to be field ionized.
This spurious contribution will add to the correct probability of ionization
for ionic impact.
As an example, in Cohen's model (Cohen 1985), with a field strength of $F_0/2$,
about 30 per cent of the electrons would have an energy beyond the ionization
threshold because of the choice of the initial conditions.
\par
Initial position and momentum have been chosen as if the electron
were not under the influence of the field. In Fig. 1 we report the evolution
of spatial distribution; it is clear that already after a short period,
$\leq 5$ au, the spatial distribution has reached the equilibrium value
compatible with the field.
\par
Target and projectile nuclei have been set at an initial distance of $20 Z$
au,
where $Z$ is the projectile charge. The equations of motion have been
integrated until the nuclei were well far apart.
For each test a number of runs ranging between $3 \cdot 10^4$ and
$ 26 \cdot 10^4$ was performed.
We did not attempt to minimize the statistical error for every point, however
for all of the data the error goes from  1 to  a
maximum of $20$ per cent.
\par
When considering the final state of the electron it is essential to remember
that an electron captured in an high enough quantum number may be
subsequently field ionized. We have followed the convention of Olson and
MacKellar (1981) of considering as ionized such electrons.
\par
Three different field geometries have been considered:  adopting the
system of reference where the neutral atom is initially at rest and the ion
is moving along the $z$ axis towards the positive direction, the electric
field has been chosen either parallel to $z$, antiparallel or perpendicular.
In the following we will adopt the convention of designate a parallel or
antiparallel field according to its projection on the $z$ axis, so
$F > 0$ means ${\vec F} \parallel {\vec v_p}$.

\vskip 0.5 truecm

{\bf III. Hydrogen--proton collisions}
\vskip 0.3 truecm
\par
The first series of calculations was about $H-H^+$ collisions: in Fig. 2
we plotted electron capture cross section, $\sigma_{cx}$, and ionization,
$\sigma_{ion}$, versus impact velocity for two different values of the field,
chosen parallel. It may be stated, alike shown in Olson and MacKellar (1981),
that $F$ enhances ionization, but at the same time opposes
to the process of recombination on the projectile,
working against charge exchange. The overall effect is
quite clearly increasing with the strength of $F$ as far as $\sigma_{ion}$ is
concerned, while for $\sigma_{cx}$ the effect is less marked, even though
visible. We note further that the field is effective on electron capture
only at low energies, its influence being negligible above $v = 2$ au
(i.e. $ E = 100$ KeV/amu). Instead, $\sigma_{ion}$ is affected in the
same manner along all the energy range.
\par
In Fig. 3 the same calculations have been performed by varying the field
direction. We may state that $\sigma_{cx}$ is dependent from the sign of $F$,
the difference being small but regular. In Olson and MacKellar (1981) it was
already suggested that when the field pulled the electron in the direction
of the outgoing projectile, electron capture and loss were augmented.
A less definite effect is visible on $ \sigma_{ion}$.
There are not appreciable differences when the field is set orthogonal to the
ion direction or parallel to it.
\par
The calculations were then repeated with a change in the initial condition,
the electron being now set in the $ n = 2$ state. From Fig. 4, where we
have plotted the data corresponding to two opposite values of the field,
 one argues that the difference between $\sigma_{cx}(F <0) $ and
$\sigma_{cx}(F>0)$ is clearly enhanced, while not a clear behavior along
the entire energy range appears for $\sigma_{ion}$.

It is well known that the, in field--free conditions, the
collision process has certain symmetries, basing upon which some scaling
laws can be inferred (see, for example, Janev 1991): all collisions between an
hydrogen atom in the state $n$ and an ion of charge $Z$ lie on the same
curve when one plots cross sections in terms of the reduced quantities
\beq
\tilde{\sigma} (\tilde{E}) = { \sigma (E) \over n^4 Z} , \;\;\;\;
\tilde{E} = { n^2 E \over \sqrt{Z} }.
\eeq
Under non zero--field conditions the same relations should hold, provided
the external field is scaled according
\beq
\tilde{F}= { F \over n^4}.
\eeq
In Fig. 5 we have plotted the same quantities of Figs. 3 and 4 in terms
of these new reduced variables; both for $\sigma_{cx}$ and $\sigma_{ion}$
a single trend is discernible, but the spreading of the points is not
negligible.
\par
As far as differential cross sections are concerned, we have plotted in
Fig. 6 the quantities $d\sigma / db$ for two scatterings at the same
impact velocity and opposite values of $F$. The
effect of $F$, in both cases,  is to reduce the effective range within which
 capture takes place, while ionization is now made possible at larger impact
parameters.

\vskip 0.5truecm

{\bf IV. Hydrogen--multiply charged ions scattering}
\vskip 0.3 truecm
\par
In order to study features of non--resonant scattering
we made simulations of collisions of hydrogen with $He^{2+}$,
$Li^{3+}$, $C^{6+}$, and $O^{8+}$ ions, whose results are reported
in Fig. 7. We may observe that a different behavior exist between the lightest
ions ($H, He$), and the others. This is also clearly revealed in Fig. 8 where
the same quantities have been plotted in terms of the reduced variables of
eq. 3; however we observe that $Z$--scaling of eq. 3 still holds for the
heaviest ions and on the whole this scaling law is obeyed quite well.
\par
A further subject is the distribution of $\sigma_{cx}$ over quantum
states of the projectile. In this work we only considered distributions
over the principal quantum number $n$. Classically, a bound electron may
have any value of the energy, below the ionization potential. Capture into
discrete energy levels is approximated by a procedure
developed, for example, in Olson (1981) and Salop (1979).
It is well known that there exist a proportionality relation between the
ion charge  $ Z$ and the quantum state $ n$ into which the electron is
preferentially captured, $ n \approx Z^{3/4} $ (Olson 1981).
In Fig. 9  we plotted these distributions for $ C^{6+}$ and
$O^{8+}$ ions: $\vec F$ has no effect when capture takes place in low
lying states while, as one could foresee, capture in higher states, which
are less bound to the nucleus, is strongly
suppressed. The position of the maximum is unaltered with
respect to the null-field case (Olson 1981, Salop 1979).
We remark that $n = 7$ and $n = 9$ are,
respectively for carbon and oxygen, the maximum non field--ionized
quantum numbers at this value of $F$.

\vskip 0.5 truecm

{\bf V. Shrinking the extension of the field}
\vskip 0.3 truecm
\par
Until now all the calculations have been performed under the hypothesis of
an uniform external field, extending in all directions up to infinity. As a
final issue we modified this configuration by limiting the field within a
finite interval: it is abruptly set to zero outside a distance $L$
 from the target nucleus along the $z-$axis in  both directions. After
this modification we have  a more symmetrical situation, in which both
Coulomb interparticle forces and $\vec F$ can exercise an influence over
only a limited region. Ideally, the length $L$  should be taken great enough
to include most of the interparticle interaction within it:
we chose $L = 10~Z$ au .
Repeating previous calculations for $H-H^{+}$ collisions gives some
modifications:
$\sigma_{cx}$ is no longer uniformly
suppressed and may become even greater than in the zero field case, and
the opposite is true for $\sigma_{ion}$ (Fig. 10). Another non negligible
difference stems when dealing with high charged ions: in Fig. 11 we have
plotted $n$--distributions for $H-C^{6+}$ impacts at a velocity
$ v_p = 2$. The absolute values of
$\sigma_{cx}$ corresponding to $F = 0, +0.03, -0.03$ au are, respectively,
$9.5 \cdot 10^{-16}$ cm$^2$, $5.4 \cdot 10^{-16}$ cm$^2$,
$17.4 \cdot 10^{-16}$ cm$^2$, and the auxiliary contribution to the
$F<0$ case comes from electrons which are bound in high lying states
around $n = 10,11$.
We suggest that this result may be explained by a mechanism similar to
capture--via--ionization (see McCartney and Crothers 1994):
the incoming projectile strikes the
electron, raising it in the continuum with an initial velocity $\vec v_0$,
which we will assume to be parallel to the ion velocity $\vec v_p$.
The electron is then accelerated by the field. On $z = L$, it reaches the
maximum speed which, with the position $ F = - |F|$, may be written
$v_f \approx \sqrt{v_0^2+2 L |F|}$.
The relative distance between the ion and the electron is now
\beq
D = v_p{ v_f -  v_0 \over |F| } - L.
\eeq
Capture takes place with preference when $v_p$ and $v_f$ closely match,
$ v_p \approx v_f$.
The binding energy is then
\beq
E \approx - { Z \over D } = - { Z^2 \over 2 n^2 } \; ,
\eeq
from which one gets
$ n \approx [\sqrt{Z D /2}] $ ([$\;$] means the integer part).
The condition $ v_p = v_f $ gives $v_0 = \sqrt{v_p^2 - 2 L  |F|}$.
The substitutions $v_p = 2, \; Z=6, \; L=60, \; |F|=0.03$ give
$ n \approx 10$ in good accordance with Fig. 11. A similar, but much
weaker, feature appears in the $ F > 0$ case; the maximum at $n = 7$ may
be partially explained by the same mechanism, provided that now the field
has a braking effect on the electron, whose initial speed must thus be greater
than $v_p$.

\vskip 0.5 truecm
{\bf VI. Summary }
\vskip 0.3 truecm
\par
In this work we have performed some numerical simulations with the aim to
investigate how a strong static electric field may modify an energetic
collision
between an ion and an atom. We studied total and partial cross sections for
charge--exchange and ionization versus some meaningful parameters as
impact velocity $v_p$, ion charge $Z$, initial target quantum state $n$ and
spatial extension of the field.
It has been found that the field deeply affects the results of the scattering;
detailed effects depend upon the strength of the field and its orientation with
respect the relative nuclear motion. Scaling laws, true in
null-field situations, seem now more doubtful, and are surely true only
within strict ranges of the parameters.

\vskip 0.5 truecm
\begin{center}
{\bf Acknowledgments}
\end{center}
\vskip 0.3 truecm
\par
L.S. acknowledges Prof. J.M.G. Gomez for his kindly
hospitality to the Department of Atomic, Molecular and Nuclear Physics
of "Complutense" University, and the "Istituto Veneto di Scienze,
Lettere ed Arti" for a partial support.
F.S. has been supported by a grant of the Ministero per l'Universit\`a
e la Ricerca Scientifica e Tecnologica; he also acknowledges the hospitality
offered by the Istituto Gas Ionizzati del CNR of Padua.

\newpage

\parindent = 0.pt

\section*{References}
\vspace{0.6 cm}

Abrines R A and Percival I C 1966a {\it Proc. Phys. Soc. }{\bf 88} 862

----- 1966b {\it Proc. Phys. Soc. }{\bf 88} 873

Bezchastnov V G and Potekhin A J 1994 {\it J. Phys. B: At. Mol. Opt. Phys. }
{\bf 27} 3349

Buchleitner A, Gremaud B and Delande D 1994 {\it J. Phys. B: At. Mol.
Opt. Phys. }{\bf 27} 2663

Bylicki M, Themelis S I  and Nicolaides C A 1994 {\it J. Phys. B: At. Mol.
Opt. Phys. }{\bf 27} 2741

Cohen J S 1985 {\it J. Phys. B: At. Mol. Phys. }{\bf 18} 1759

Damburg R J and Kolosov V V 1976 {\it J. Phys. B: At. Mol. Phys. }{\bf 9} 3149

Delande D and Gay J C 1986 {\it Phys. Rev. Lett. }{\bf 57} 2006

Delande D {\it et al} 1994 {\it J. Phys. B: At. Mol. Opt. Phys. }{\bf 27} 2771

Eichenauer D, Gr\"un N and Scheid W 1981
{\it J. Phys. B: At. Mol. Phys. }{\bf 14} 3929

Eliezer S {\it et al} 1988 Laser Interaction and Related Plasma
Phenomena (eds. Hora H and Miley P G, Plenum, New York, vol. 8)
p. 279

Eliezer S and Hora H 1989 {\it Phys. Rep. }{\bf 172} 339

Friedrich H 1990 Atoms in Strong Fields (eds. C.A. Nicolaides, C. Clark
and M. Nayfeh) p. 247

Grosdanov T P and McDowell M R C 1985 {\it J. Phys. B: At. Mol. Phys. }
{\bf 18} 921

Hardie D J W and Olson R E 1983 {\it J. Phys. B: At. Mol. Phys. }{\bf 16} 1983

Janev R K 1991 {\it Phys. Lett. A} {\bf 160 } 67

Kunc J A 1988 {\it J. Phys. B: At. Mol. Phys.} {\bf 21} 3619

McCartney M and Crothers D S F 1994 {\it J. Phys. B: At. Mol. Opt. Phys.}
{\bf 27} L485

McDowell M R C and Zardona M 1985 {\it Adv. At. Mol. Phys. }{\bf 21} 255

Olson R E 1981 {\it Phys. Rev.} A {\bf 24} 1726

Olson R E and MacKellar A D 1981 {\it Phys. Rev. Lett. }{\bf 46} 1451

Raadu M A 1989 {\it Phys. Rep. }{\bf 178} 25

Reinhold C O and Falcon C A 1988a {\it J. Phys. B: At. Mol. Phys. }
{\bf 21} 1829

----- 1988b {\it J. Phys. B: At. Mol. Phys. }{\bf 21} 2473

Ruder H, Wunner G, Herold H and Geyer F 1994
Atoms in Strong Magnetic Fields (Springer--Verlag)

Salop A 1979 {\it J. Phys. B: At. Mol. Phys.} {\bf 12} 919

Schmidt A, Horbatsch M and Dreizler R M 1990
{\it J. Phys. B: At. Mol. Phys. }{\bf 23} 2327

Seipp I and Taylor K T 1994 {\it J. Phys. B: At. Mol. Opt. Phys. }{\bf 27} 2785

Wang Q and Greene C H 1991 {\it Phys. Rev.} A {\bf 44} 7448

Williams A C {\it et al} 1986 {\it Astrophys. J.} {\bf 305} 759

Zang J X and Rustgi M L 1994 {\it Phys. Rev.} A {\bf 50} 861

\newpage

\section*{Figure Captions}
\vspace{0.6 cm}

{\bf Figure 1}: upper, radial distribution; lower, polar angle distribution.
Full line represents microcanonical distribution at $F = 0$.
We set up the field $F = 0.03$ au at $t = 0.0$.
Our plots show distributions at $ t = 0.0$ (full circles),
$t = 5.0$ au (empty circles), $t = 10.0$ au (full squares).
 A total of $10^5$ runs have been performed.

{\bf Figure 2}: upper, $ \sigma_{cx}$; lower $ \sigma_{ion} $ versus impact
velocity. Full diamonds, $ F = 0.0$; open circles, $ F = 0.03$ au ;
 full circles, $F = 0.04$ au .

{\bf Figure 3}: upper, $ \sigma_{cx}$; lower $ \sigma_{ion} $ versus impact
velocity.
Full diamonds, $ F = 0.0$; open circles, $ F = 0.03$ au; full circles,
$F = -0.03$ au; open squares, $ F = 0.03$ au (orthogonal).

{\bf Figure 4}: upper, $ \sigma_{cx}$; lower $ \sigma_{ion} $ versus impact
velocity. Open circles, $ F = 0.03/2^4$ au;
full circles, $ F = -0.03/2^4$ au. The electron is in
the initial status $ n = 2$.

{\bf Figure 5}: upper, charge exchange; lower, ionization. In ordinate appear
scaled cross sections $  \sigma/n^4$; in abscissae
scaled velocity $  v~n $. Open circles, $ F = 0.03$ au, $n = 1$;
full circles, $ F = 0.03/2^4$ au, $n = 2$; open squares,
 $ F = -0.03$ au, $n = 1$;
full squares, $ F = - 0.03/2^4$  au, $n = 2$.

{\bf Figure 6}: upper, $ { d\sigma_{cx}/db}$;
lower, ${d\sigma_{ion}/db}$
versus impact parameter $b$. Impact energy is kept fixed at $ E = 50$ KeV/amu.
Full diamonds, $ F = 0.0$;
open circles, $ F = 0.03$ au; full circles, $ F = -0.03$ au.

{\bf Figure 7}: upper, $\sigma_{cx}$;
lower $ \sigma_{ion} $ versus impact velocity.
Field strength is kept fixed at 0.03 au. Full triangles $H^+$;
full circles $ He^{2+}$; open circles $Li^{3+}$; full squares $C^{6+}$.

{\bf Figure 8}: upper, charge exchange; lower, ionization. In ordinate appear
scaled cross sections $ \sigma/Z$; in abscissae scaled velocity
$ v/Z^{1/4} $. Open triangles, $ O^{8+}$; the other symbols are the
same as in Figure 7.

{\bf Figure 9}: distribution of $\sigma_{cx}$ over $n$. Upper, $C^{6+}$;
lower, $O^{8+}$.
Open squares $ E = 50$ Kev/amu; full squares $ E = 100$ keV/amu.

{\bf Figure 10}: upper, $\sigma_{cx}$;
lower $\sigma_{ion}$ versus impact velocity. Full diamonds, $F = 0.0$;
open circles, $ F = 0.03$ au; full circles, $ F = -0.03$ au;
open squares, $ F = 0.03$ au (orthogonal). The length $L$  has
been kept fixed at $ L = 10$ au.

{\bf Figure 11}: distribution of $\sigma_{cx}$ over $n$
for $C^{6+}-H$ impacts at an energy $ E = 100$ KeV/amu.
Diamonds, $F = 0.0$; open circles, $F =  0.03$ au; full circles,
 $F= -0.03$ au. The length $L$ is kept fixed, $ L = 60$ au.

\end{document}